\documentclass[12pt,twoside]{article}   
\usepackage[super,sort,comma]{natbib}

\usepackage{fancyhdr}		

\usepackage[section]{placeins}   %

\usepackage{graphicx}

\makeatletter \renewcommand\@biblabel[1]{$^{#1}$} \makeatother
\newcommand{\cen}[1]{\begin{center} #1 \end{center}}

\usepackage{hyperref}
\hypersetup{ colorlinks,
    citecolor=blue,
    filecolor=blue,
    linkcolor=blue,
    urlcolor=blue
}

\usepackage{xcolor}

\definecolor{gray}{rgb}{0.6,0.6,0.6}
\definecolor{red}{rgb}{0.85,0,0}
\definecolor{green}{rgb}{0,0.85,0}
\definecolor{blue}{rgb}{0,0,0.85}
\definecolor{beige}{rgb}{0.92,0.87,0.78}
\usepackage[all]{hypcap}    

\usepackage{amsmath}
\newcommand\norm[1]{\lVert#1\rVert}

\usepackage{amssymb}
\newcommand\numberthis{\addtocounter{equation}{1}\tag{\theequation}}

\DeclareMathOperator*{\argmin}{arg\,min}

\begin{document}

\cen{\sf {\Large {\bfseries Surface guided analysis of breast changes during post-operative radiotherapy by using a functional map framework } \\  
\vspace*{10mm}
Pierre Galmiche$^a$, Hyewon Seo$^a$, Yvan Pin$^b$, Philippe Meyer$^{a,c}$, Georges Noël$^{a,c}$, Michel De Mathelin$^a$} \\
$^a$ Laboiratoire ICube, CNRS--University of Strasbourg, Strasbourg, France\\
$^b$ Institut Privé de Radiothérapie de Metz (IPRM)\\
$^c$ Institut de Cancérologie de Strasbourg Europe(ICANS)
\vspace{5mm}\\
Version typeset \today\\
}

\pagenumbering{roman}
\setcounter{page}{1}
\pagestyle{plain} Corresponding author: Hyewon Seo,  2 Rue Marie Hamm, 67000 Strasbourg, France. email: seo@unistra.fr \\
\begin{abstract}
\noindent {\bf Background:} The treatment of breast cancer using radiotherapy involves uncertainties regarding breast positioning. As the studies progress, more is known about the expected breast positioning errors, which are taken into account in the Planning Target Volume (PTV) in the form of the margin around the clinical target volume. However, little is known about the non-rigid deformations of the breast in the course of radiotherapy, which is a non-negligible factor to the treatment. \\ 
{\bf Purpose:} Taking into account such inter-fractional breast deformations would help develop a promising future direction, such as patient-specific adjustable irradiation plannings. \\
{\bf Methods:} In this study, we develop a geometric approach to analyze inter-fractional breast deformation throughout the radiotherapy treatment. Our data consists of 3D surface scans of patients acquired during radiotherapy sessions using a handheld scanner. We adapt functional map framework to compute inter- and intra-patient non-rigid correspondences, which are then used to analyze intra-patient changes and inter-patient variability. \\
{\bf Results:}The qualitative shape collection analysis highlight deformations in the contra-lateral breast and armpit areas, along with positioning shifts on the head or abdominal regions. We also perform extrinsic analysis, where we align surface acquisitions of the treated breast with the CT-derived skin surface to assess displacements and volume changes in the treated area. On average, displacements within the treated breast exhibit amplitudes of 1–2 mm across sessions, with higher values observed at the time of the 25\textsuperscript{th} irradiation session. Volume changes, inferred from surface variations, reached up to 10\%, with values ranging between 2\% and 5\% over the course of treatment. \\
{\bf Conclusions:} We propose a comprehensive workflow for analyzing and modeling breast deformations during radiotherapy using surface acquisitions, incorporating a novel inter-collection shape matching approach to model shape variability within a shared space across multiple patient shape collections. We validate our method using 3D surface data acquired from patients during External Beam Radiotherapy (EBRT) sessions, demonstrating its effectiveness.

The clinical trial data used in this paper is registered under the ClinicalTrials.gov ID NCT03801850.
\end{abstract}


\tableofcontents

\newpage

\setlength{\baselineskip}{0.7cm}      

\pagenumbering{arabic}
\setcounter{page}{1}
\pagestyle{fancy}
\section{Introduction}

As a mobile external organ, the mammary gland is prone to various uncertainties regarding its daily positioning. Moreover, regular clinical examinations of patients during the post-operative breast radiotherapy show frequent cases of progressive changes in breast morphology. 
These changes may be attributed to both systematic and stochastic daily factors, including inter-fraction positioning errors (such as arm elevation required to avoid irradiation during treatment), respiratory motion, seroma involution\cite{sharma_change_2009}, and oedema/inflammation phenomena.\cite{franco_tomodirect_2011,mohiuddin_decrease_2012,oh_planning_2006}
Indeed, a regular clinical examination of patients during the irradiation treatment showed inter- and intra-fractional breast changes observed the during radiotherapy sessions.
T. Alderliesten et. al. \cite{alderliesten_breast-shape_2018} estimated inter-fractional breast variations in the order of 4mm. J. Seppälä et. al. \cite{seppala_breast_2019} observed a maximum breast surface expansion up to 15mm with 17\% superior to 5mm, suggesting the need for an additional 8 mm of margin to cover the whole breast in 95\% of the treated fractions in their study.
In R. Ricotti et. al.\cite{ricotti_intra-fraction_2017}, the average intra-fractional respiratory motion of the chest in the antero-posterior direction has been evaluated to be less than 2mm on average.

While the exact mechanisms behind these morphological changes remain uncertain, a systematic analysis of alterations in breast shape and volume is essential for assessing their subsequent impact on treatment quality.
In this paper, we develop a method for measuring and analyzing the irradiated breast shape and volume evolution during the radiotherapy following breast conservative surgery. Based on a surface dataset acquired from more than 20 patients during radiotherapy session,
our developed method enables the collective analysis of geometric deformation during the course of a radiotherapy treatment. 
The proposed method performs the following steps: (1) Computation of correspondences between the acquisitions of a patient, (2) Rigid alignment of the surfaces to the CT, (3) inter- and intra-patient variability analysis and volume changes estimation.
Our work aligns with these studies as we aim to investigate such inter-fractional shape changes.
However, it stands out as the first methodology to geometrically analyze collections of breast radiotherapy surface data using functional maps.

Surface data is increasingly being used for patient positioning. The positioning of the breast/patient is a crucial aspect of radiotherapy because it directly impacts the delivery of the treatment dose. Ensuring the patient maintains the same position as during the centering CT scan, which was used to plan the treatment, is essential. Significant changes in the patient's position can result in the delivered radiation differing greatly from the planned dose, potentially causing secondary effects due to over-dosage.
To make sure the patient is well positioned, recent approaches tend to use surface acquisitions. Indeed, older approaches such that the use of bones structures\cite{padilla_assessment_2014} have shown limitations, due to the large non-rigid deformations of the women torso.
Recent surface-guided radiotherapy (SGRT) utilizes frequently measured surface data during treatment to prevent irradiation when patient positioning errors are unacceptable.
Despite its advantages over the older laser-based setup in terms of cost, accuracy, setup speed and patient comfort \cite{kugele_surface_2019,gonzalez-sanchis_surface-guided_2021, penninkhof_evaluation_2022}, this approach is mainly used for measuring the positioning error on the surface level and does not exploit the non-rigid deformations captured in the various acquisitions.
Surface data acquired during radiotherapy treatment are highly valuable and have recently been the focus of research aimed at enhancing treatment effectiveness. A. Gorecki et. al. \cite{Gorecki23} proposed an approach to modify the CT volume so that it takes into account the SGRT surface deformations. In the future, advances of this kind could be used to adapt dose maps based on surface deformations.
Our approach has been evaluated using surface data collected before, during, and after treatment in a clinical trial conducted by the Institute of Cancerology Strasbourg Europe (ICANS). Notably, our approach is in line with P. Freisledere et. al.\cite{freislederer_recent_2020}, which recommended using deformable surface registration to create more precise patient motion tracking methods.

The main contributions of the work are thus three-fold:
\begin{itemize}

\item We have acquired  patients’ 3D surface data during External Beam Radiotherapy (EBRT) sessions, on which we validate our method. 
    
\item We propose a complete workflow for analyzing and modeling radiotherapy breast deformations from surface acquisitions. 

\item We introduce a new inter-collection shape matching approach, allowing to model shape variability in a space common to various patient shape collections.

\end{itemize}

\noindent To the best of our knowledge, our work is the first computational approach to address the problem of non-rigid breast deformation during radiotherapy with functional maps.

\section{Data and Methods}

\subsection{Dataset}
\label{dataset}

The patient's surface data collection protocol has been approved by the Ethics Committee for the Protection of Individuals at the Faculty of medicine of the University of Strasbourg, which involved the written informed consent from all patients in this study (No RCB: 2017-A02489-44).
To systematically study breast morphology changes throughout the radiotherapy process and during the post-treatment period, surface scans of patients undergoing post-operative radiotherapy were acquired as part of a clinical trial conducted by the Institute of Cancerology Strasbourg Europe (ICANS), registered under the ClinicalTrials.gov ID NCT03801850.
  \begin{figure}[htpb]
        \centering
        \includegraphics[width=\linewidth]{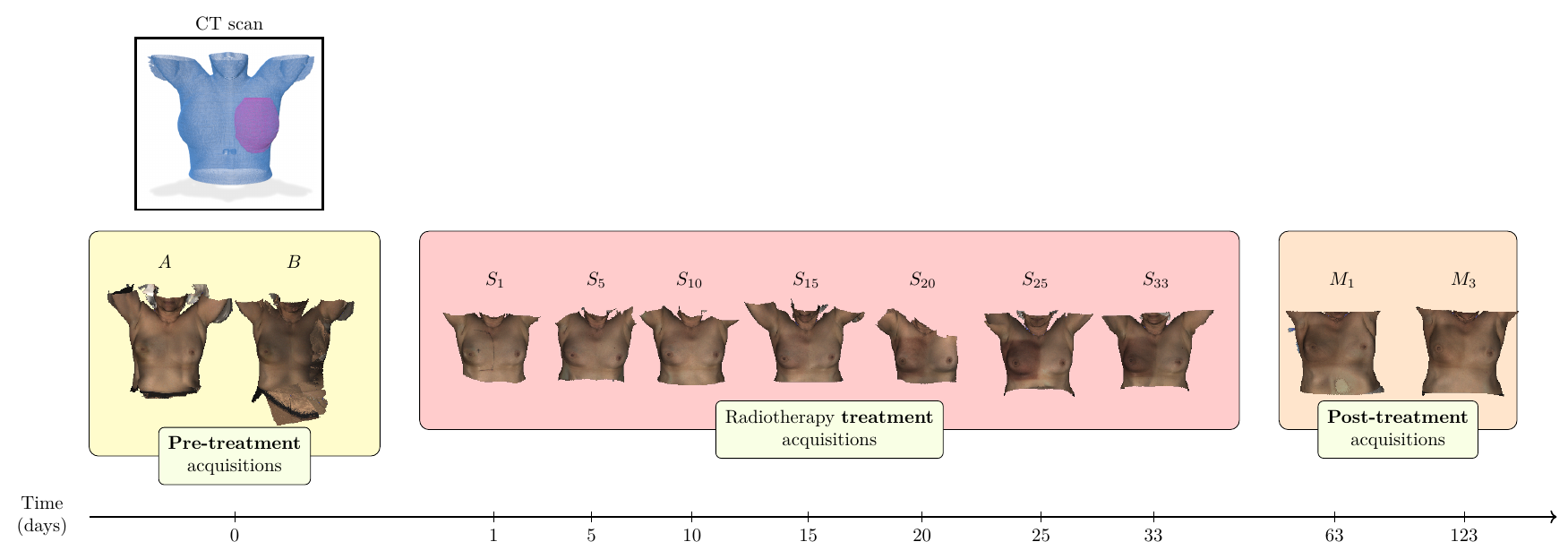}
\caption{Illustration of our clinical trial data acquired for each patient. Optical scans were captured as textured 3D meshes during the trial (shown below). 
The surfaces $A$ and $B$ have been acquired on the same day as the planning $CT$ scan, while $S_X$ has been acquired during the $X^{th}$ irradiation session. Finally $M_1$ and $M_3$ have been acquired one month and three months after the last irradiation. In addition to the clinical trial surfaces, we also consider the point clouds representing the patient skin (blue) and treated breast (pink), annotated from the CT scan (shown at the top).
}
        \label{fig:ICANSdata}
\end{figure}

The trial involved 60 participants who had undergone breast-conserving surgery and were receiving post-operative radiotherapy. Each patient underwent a standard course of 25 sessions of normofractionated breast radiotherapy, delivering 2 Gy per session, followed by a boost dose administered over 8 additional sessions of 2 Gy.
In addition to the standard planning procedures, which included planning CT scans and the identification of key structures (e.g., treated breast, lungs, and skin) for therapy planning, multiple surface scans were performed by using a hand-held optical scanner\cite{artecEva} to monitor the inter-fraction evolution of the breast.
Before starting an irradiation session, the operator swept the hand-held scanner slowly over the torso while paying attention to the coverage of the breasts.

Two optical surface scans were acquired right after the planning CT scan, to evaluate the repeatability of the measurement data. Then, during the treatment period, an optical surface scans was acquired every five sessions, resulting in 5 to 7 surface data.
Finally, two additional acquisitions were taken one month and three months after the end of the treatment to observe post-treatment changes.
The whole dataset contains 23 patients with 8 to 11 surface scans per patient, making a total of 247 surface acquisitions available for this study.

Figure \ref{fig:ICANSdata} summarizes the dataset we collected from each participant in our study.
Depending on the time of acquisition, they are labeled either as \textit{pre-treatement}, \textit{treatement}, or \textit{post-treatement}.
Each patient data is composed of two types of data:
\begin{itemize}
    \item CT skin mesh with the treated breast region segmented: We generate the mesh from a point cloud (Figure\ref{fig:ICANSdata}(a)) derived from the \textit{planning CT scan} images, where the organs of interest are annotated by clinicians.    
 
    \item Surface meshes: Triangular meshes with about 10,000 points (Fig. \ref{fig:ICANSdata}(c)), after the simplification of initial mesh acquired by the hand-held scanner.
    They represent different coverages of the patients' torso, are not rigidly aligned, and contain noise. 
\end{itemize}

\subsection{Overview}
\begin{figure*}[htpb!]
    \centering
    \includegraphics[width=0.95\linewidth,height=0.8\textheight, keepaspectratio]{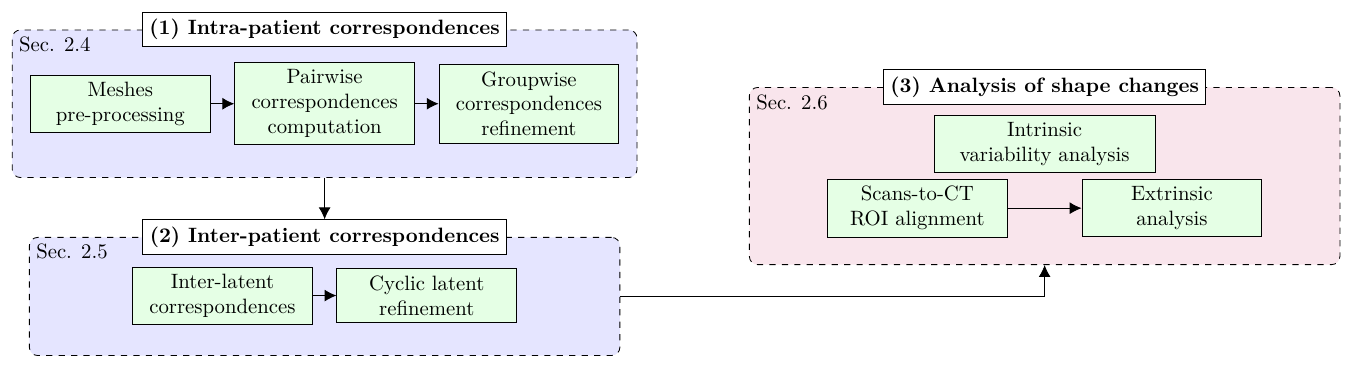}
    \caption{Schematic illustration of our approach and its components. From the clinical dataset, we first compute correspondences between the scans of the same patient focusing on the ROI (1). From these intra-patient correspondences, and derive efficiently inter-patient correspondences (2) that are used for the breast shape evolution throughout radiotherapy (3).}
    \label{fig:intrapatient}
\end{figure*}

To model the evolution of the women's bust during radiotherapy, we propose a complete pipeline with three major steps:
\begin{itemize}
    \item First, we compute \textit{intra-patient} correspondences between the acquisitions of single patient and use them to establish \textit{inter-patient} correspondences (Figure \ref{fig:intrapatient} (1) and (2)).
    \item We then rigidly align all the optical acquisitions of the patient to their respective CT skin using the intra-patient correspondences from the previous step.
    \item Finally, we analyze the shape variations in our dataset both quantitatively and qualitatively (Figure \ref{fig:intrapatient}(3) and (4)).
\end{itemize}

The proposed intra-patient correspondences pipeline is described in Section \ref{intra-patient-corr}. We first down-sample and clean each surface by removing noise, smoothing, and ensuring that the mesh consists of a single connected component.
Then, we initialize pairwise correspondences between the \textit{CT skin} and each optical scan of the same patient, followed by \textit{scan-to-scan} correspondences between consecutive optical scan acquisitions. The vertex-to-point correspondence computation is performed by refining the Nearest Neighbors correspondences using the ZoomOut approach \cite{melzi_zoomout_2019}.
We then apply the shape collection refinement method based on cycle-consistency from P. Galmiche et al.\cite{Galmiche_Seo_2023} to compute the final intra-patient non-rigid correspondences, with a focus on the annotated breast region as the Region on Interest (ROI).

With the ensemble of intra-subject correspondences, we then compute inter-patient correspondences, establishing the correspondence to all meshes in the dataset (Section \ref{inter-patient-corr}). This allows us to computationally analyze breast shape changes during radiotherapy. In Section \ref{deformation_modeling}, we describe our methods and present results on intrinsic and extrinsic shape analysis across this collection of breast shapes.

\subsection{Preliminaries}
\label{preliminaries}
\noindent \textbf{Spectral mesh representation.} We assume a collection of $N$ 3D shapes $\mathcal{S}^{id} = \{S_n^{id}\}_{n=1}^N$ for each patient $P^{id}$, represented as triangular meshes.
To each shape $S_n^{id}$, we associate a positive semi-definite Laplacian matrix $L_{n} = A_{n}^{-1} W_{n}$ by using the standard cotangent weight scheme of U. Pinkall et. al.\cite{pinkall_computing_1993}, where $ W_{n}$ is the cotangent weight matrix and $A_{n}$ is the diagonal lumped area matrix.
The eigendecomposition of $L_n$ form the spectral representation of the shape. Specifically,
the basis functions $\Phi^{S_n} = \{ \phi_j^{S_n} \}_{j \geq 1}$ composed of the eigenfunctions of $L_n$ form an orthonormal basis of the square integrable functions on $S_n$, denoted as $\mathcal{L}^2(S_n)$, allowing to represent any function $f \in \mathcal{L}^2(S_n)$ using a Fourier series: $f = \sum_{j \geq 1} \langle f, \phi_j^{S_n} \rangle \phi_j^{S_n}$.
We use this spectral representation of the shapes for its compactness and flexibility, as showcased by the functional map framework which we describe below.

\vspace{0.2cm}
\noindent\textbf{Functional correspondences.} The functional map framework originally presented by M. Ovsjanikov et. al.\cite{ovsjanikov_functional_2012} aims at finding correspondences between shapes $S_i$ and $S_j$ as a linear transformation $\tau: L^2(S_i) \to L^2(S_j)$ between their respective function spaces. As a linear operator, $\tau$ admits a matrix representation $C=(c_{kl})$ with the coefficient determined as follows. The action of $\tau$ on a function $f \in L^2(S_i)$ can be expressed as :

\begin{align*}
    \tau f & = \tau \sum_{k \geq 1} \langle f, \phi^{S_i}_k \rangle \phi^{S_i}_k
     = \sum_{k\geq 1} \langle f, \phi^{S_i}_k \rangle \tau \phi^{S_i}_k \\
       & =\sum_{k,l \geq 1} \langle f, \phi^{S_i}_k \rangle_{S_i} \underbrace{\langle \tau \phi^{S_i}_k , \phi^{S_j}_l \rangle_{S_j}}_{c_{kl}} \phi^{S_j}_l. \numberthis \label{eqn}
\end{align*}

Truncating the Fourier series to the first $K$ terms allows to obtain a particularly compact representation of the functional correspondences with a $K \times K$ matrix $C$, where $K$ is chosen to be small (between 20 and 100 in practice). 

A general approach to optimize such a matrix assumes the availability of corresponding functions $g_c \approx \tau f_c, ~ c = 1, \ldots, Q$ and asks the functional map $C$ to conserve the spectral coefficients in a least square sense: 
\begin{equation}
    \min_C \norm{CA-B}^2_F + \alpha E_{reg}(C), 
\end{equation}
where \textbf{A}$=(\langle \phi^{S_i}_k,f_c\rangle)$
and \textbf{B}$=(\langle \phi^{S_j}_l,g_c\rangle)$ are $K \times Q$ matrices of Fourier coefficients of the corresponding functions, $\alpha$ is a scalar weight and $E_{reg}(C)$ is a regularization term.

Recovering a point-wise map from a given functional map can be expressed as the solution of the following optimization problem:
\begin{equation}
    T_{ij}(p) = \argmin_{q \in S_j} \norm{C_{ji} \Phi^{S_j}_{k}(q)^T - \Phi^{S_i}_{k}(p)^T}_2 ~, ~ \forall p \in S_i,
    \label{eq:FMsimple}
\end{equation}
which can be efficiently solved by searching for each row of $\Phi^{S_i}$ its nearest-neighbor in the space of rows of $\Phi^{S_j}$, transformed by $C_{ji}$.
Indeed, any point $p \in S_j$ can be represented by the corresponding delta function $\delta_p \in \mathcal{L}^2(S_j)$, expressed in $\Phi^{S_j}$ as $$\Phi^{S_j} (p) = \left[ \phi^{S_j}_1(p), \phi^{S_j}_2(p), \ldots, \phi^{S_j}_K(p) \right].$$
Conversely, a functional map $C_{ji}$ can be obtained from a point-to-point map $T_{ij} : S_i \to S_j$ using the following equation: 
\begin{equation}
C_{ji} = (\Phi^{S_i})^{\dag} \Pi_{ij} \Phi^{S_j},
\label{ptptofmap}
\end{equation}
where $\Pi_{ij}$ is the matrix representation of $T_{ij}$ such that \newline $\Pi_{ij}(index(p), index(q)) = 1$ if $T_{ij}(p) = q$ and 0 otherwise, with $p \in S_i$ and $q \in S_j$.

\subsection{Intra-patient correspondence}
\label{intra-patient-corr}

We start by computing the correspondences among different surface acquisitions collected from a same patient, i.e. intra-subject correspondences.
The main difficulty originates from two facts: First, they are real-world data with significant level of noises. This makes existing feature-based methods inappropriate, as they suffer from the inconsistent feature descriptors computed on the same semantic regions on different surfaces.
Second, the surface scans are partial measurements of the participant's torso captured with hand-held scanner,  each with different coverage. 
Optimizing correspondences between surfaces with varying coverage is complex, as it involves much larger solution space compared to scenarios where the shapes represent the same object in its entirety. To address this, we base our approach on a groupwise shape correspondence refinement with a focus on the breast region as ROI \cite{Galmiche_Seo_2023}.

\begin{figure}[htpb]
    \centering
    \includegraphics[width=1.0\textwidth]{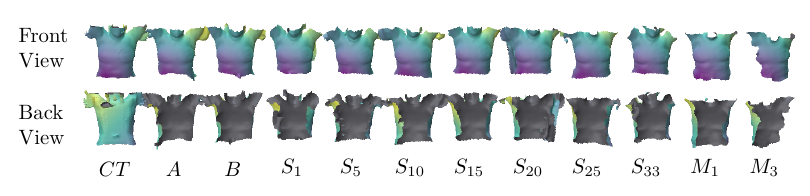} 
    \caption{An example of the scan-to-CT initial correspondences. Note that all scan meshes cover only the front part of the torso, except for the CT skin mesh which also includes the backside.}
    \label{corr_preciseNN}
\end{figure}

\vspace{0.25cm}
\noindent\textbf{Initial correspondence.} To obtain initial pairwise correspondences, we have initially applied approaches involving partial shape matching\cite{litany_fully_2017}. However, we obtained unsatisfactory results for some shape pairs containing significant noise and coverage differences.
In such cases, we manually align the scan acquisitions $A, B, S_1, \ldots, S_{33}, M_1,$ and $M_3$ to the CT, from which point-wise correspondences are defined from euclidean proximity -- For each point of the $S_i$ is considered to be in correspondence with its closest counterpart on the CT skin mesh.
As a result, we obtain initial correspondences of sufficient quality, despite the differences in coverage between the shapes. 
As shown in Figure \ref{corr_preciseNN}, correspondences thus obtained across the multiple shapes of a patient are smooth and represent the same anatomical regions, even in cases of significant coverage differences, such as between the CT and M3, where only the left side of the torso has been acquired.

\noindent\textbf{Group-wise correspondence refinement.}
These initial correspondences are subject to alignment uncertainties and errors induced by non-rigid deformations of the patient torso. To refine them by considering all the shapes of a given patient, we employ a spectral approach similar to that in P. Galmiche et. al. \cite{Galmiche_Seo_2023}.
Based on the functional map framework, it
iteratively refines functional correspondences between the patient shapes
until they are cycle consistent, i.e. the composition of maps forming cycles are close to identity. Additionally, it refines map consistency within a user-defined region of interest (ROI), which, in our case, is the treated breast. In the following, we detail 1) how we identify ROI in our data and 2) clarify the concept of cycle-consistent correspondence refinement using functional maps.

\begin{figure*}[htpb]
    \centering
    \includegraphics[width=\textwidth]{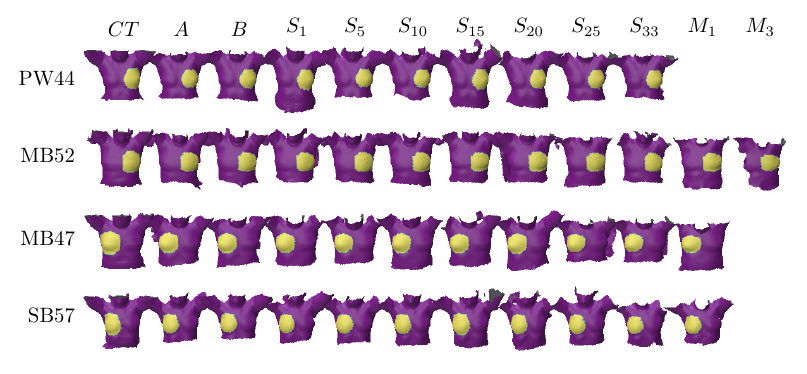} 
    \caption{ROI projected from the CT skin to the other acquisitions on four different patients undergoing right (two first rows) or left (two last rows) breast radiotherapy.}
    \label{ROI_proj_1_patient}
\end{figure*}

\vspace{0.15cm}
\noindent (1) Identification of the Region of Interest (ROI):
In our study, identifying the treated breast region or ROI on all acquisitions is a key step in analyzing breast changes.
We first locate it on the CT surface, thanks to the point cloud annotation of the treated breast provided by clinicians. 
Following this, the breast region is identified on the other scans using the correspondences between the CT and those scans. A point on a scan is considered part of the ROI if its corresponding point belongs to the ROI.
Figure \ref{ROI_proj_1_patient} shows the transferred regions obtained on four different patients, two of them with the left breast as ROI (top rows), and two other that followed right breast radiotherapy. Each scan region of interest is obtained from the cycle-consistent non-rigid correspondences.

\vspace{0.15cm}
\noindent (2) Cycle-consistency refinement:
\label{groupwiseref}
A common strategy to refine maps in a collection manner is to use cycle consistency constraints. \cite{huang_functional_2014, wang_image_2013, gao_isometric_2020, marin_spectral_2021}
The main idea is to ensure that the composition of maps, starting and ending on the same shape, results in the identity.
In the functional map framework, a shape collection is represented as a directed graph $\mathcal{G}(\mathcal{E}, \mathcal{V})$, also called a functional map network (FMN). The nodes $\{S_i\}_{i=1}^n \in \mathcal{V}$ of the graph represent the shapes of the collection in their spectral form, while an oriented edge $e_{ij} \in \mathcal{E} $  between two shapes $S_i$ and $S_j$ represent the functional map $C_{i,j}$ between the two shapes.
The cycle-consistency of a FMN $\mathcal{G}$ is defined 
as the deviation of the compositions of the functional maps along cycles from the identity. If all the maps in the collection are perfectly consistent, we have:
\begin{equation}
\begin{split}
    & C_{i_1,i_k} C_{i_k ,i_{k-1}} \ldots C_{i_2,i_1} = I, \\
    & \text{where } (i_1, i_2 \ldots i_k, i_1) \text{ is any cycle in } \mathcal{G}.
\end{split}
\label{eq:cycle}
\end{equation}
F. Wang et. al. \cite{wang_image_2013} have shown that optimizing a set of \textit{latent bases} $Y=\{Y_i\}_{i=1}^n$ such that 
\begin{equation}
    C_{ij}Y_i \simeq Y_j ~ \forall i,j
\end{equation}
 is equivalent to solving Equation (\ref{eq:cycle}), and computed latent bases Y as follows:
 
\begin{align}
    \begin{split}
      Y = & \argmin_{Y} \sum_{(i,j) \in \mathcal{E}} \omega_{ij} \norm{ C_{ij}Y_i - Y_j}_F^2, \\
     & s.t. Y^TY = I, ~ Y =  [ Y_1; Y_2; \ldots;Y_n ] \label{cycleconsisteq}
    \end{split}
\end{align}
,where $\omega_{ij}$ is a scalar reflecting the relative confidence assigned to each functional map of the network.
In our work, we utilize the \textit{Canonical Consistent Latent Bases} (CCLB)\cite{huang_limit_2019}, where an additional regularization term has been added to ensure the $\sum_i Y_i^T \Lambda_i Y_i$ is a diagonal matrix in order to overcome instability in the extracted latent bases.

\begin{figure}[htpb]
    \centering
    \includegraphics[width=0.95\linewidth,height=0.8\textheight, keepaspectratio]{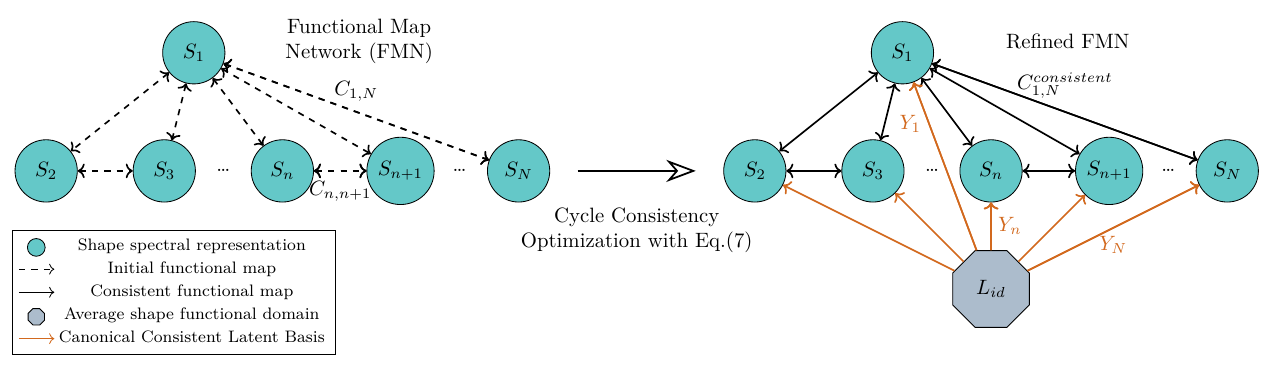}%
   \caption{Function Map Network (FMN) graph of patient $P^{id}$, before (left) and after (right) cycle-consistency optimization. The output after the functional maps consistency optimization is a latent space representing the functional domain of an average shape $L_{id}$ and its relations $Y_n$ to the other shapes of the collection.}
    \label{FMNid}
\end{figure}

The CCLB $Y=\{Y_i\}_{i=1}^N$ can be interpreted as a set of functional maps, mapping from the functional domain of a \textit{limit shape} $L_{id}$, representing the entire shape collection of patient $id$, to the individual $S_i$ in the collection. 
Figure \ref{FMNid} illustrates the FMN (left) and the refined FMN (right) after the cycle-consistency optimization 
we use for our dataset. The limit shape $L_{id}$ represent the patient $P^{id}$ entire shape collection, and are used for cross collection correspondences, as described in the next section.
Such approach has several advantages. First, there is no need to explicitly define or create a reference shape. Instead, the abstract "limit" or "average" shape $L_{id}$ is represented only in the functional domain, through the latent bases $\{Y_i\}_{i=1}^N$.
Second, we can directly recover functional maps between different acquisitions from the CCLB by using the formula $C_{ij}^{consistent} = Y_jY_i^{-1}, ~ \forall ~ i,j ~ \in 
\mathcal{E}$. 
Likewise, we can compute point-wise maps using consistent ZoomOut approach\cite{huang_consistent_2020}, which reduces to nearest neighbor searches of each row of $\Psi^{S_i} = \Phi^{S_i} Y_i$ among the rows of $ \Psi^{S_j} = \Phi^{S_j} Y_j$. This is solved using the following formulation:
\begin{equation}
    T_{ij} (p) = \argmin_{q\in S_j} \norm{\Psi^{S_j}(q)^T - \Psi^{S_i}(p)^T}_2 , ~ \forall p \in S_i.
    \label{pointwisefromcclb}
\end{equation}
This allows to recover efficiently consistent pointwise maps between any pair of shapes in the collection. 
Lastly, the construction of the latent basis has an appealing scaling property. 
Given a new, previously unseen shape $S_k$ and a functional map $C_{i,k}$ from $S_i$ to $S_k$, where $S_i$ is a shape from the collection used to estimate the latent shape, the latent basis can be pushed to $S_k$ via the equation $Y_k = C_{i,k}Y_i$.

\subsection{Inter-patient correspondences}
\label{inter-patient-corr}
Once correspondences have been estimated across all shapes for each patient, we compute correspondences between the shapes of different patients by computing
maps between the limit shapes $L_{id}$, each representing the entire collection of a patient. In order to match two limit shapes of two patients, we first pick one arbitrary shape from each patient collection: the CT mesh. We then match their limit shapes using an existing spectral matching
technique.

\begin{figure*}[htpb!]
    \centering
    \includegraphics[width=0.95\linewidth,height=0.8\textheight, keepaspectratio]{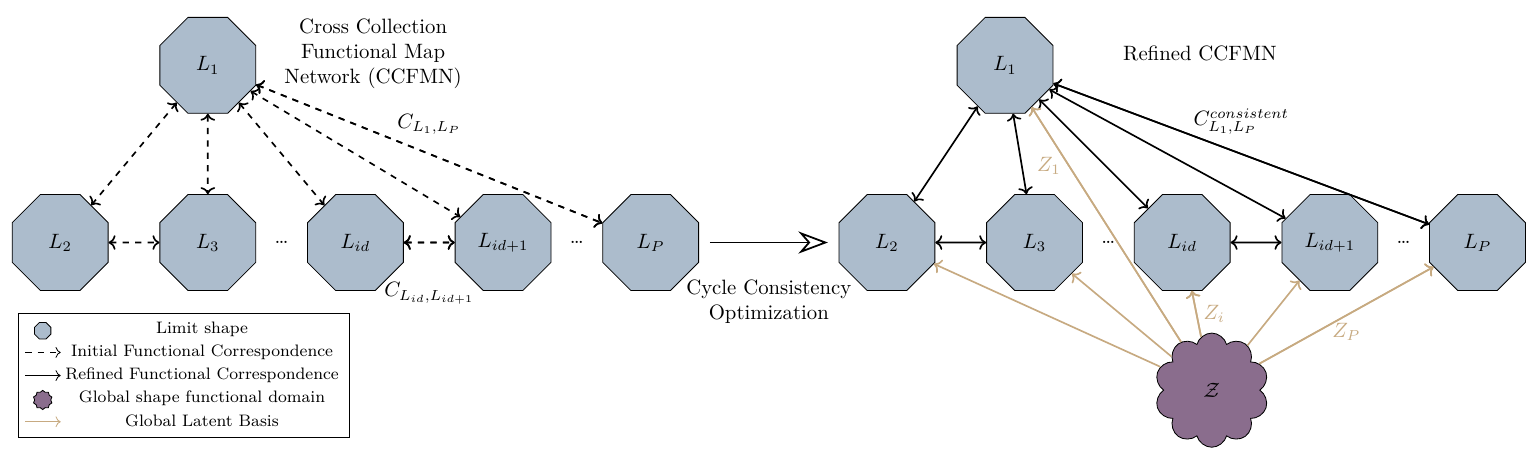}%
   \caption{Cross collection Function Map Network (CCFMN) graph, before (left) and after (right) cycle-consistency optimization. The output after the functional maps consistency optimization is a latent space $\mathcal{Z}$ representing the functional domain of a global shape and refined functional maps between the average shape of each collection.}
    \label{CCFMN}
\end{figure*}

By repeating this process between multiple pairs of patient collections, we build inter-patient shape correspondences, represented as functional maps between their respective limit shapes. 
We propose to represent all the patients in a \textit{Cross Collection Functional Map Network} (CCFMN), where the nodes of the graph are the limit shapes $\{L_{id}\}_{id=1}^P$ obtained from the intra-patient correspondences refinement, and the edges are the functional maps between the limit shapes (Figure \ref{CCFMN} (left)). From there, we construct a \textit{global shape} $\mathcal{Z}$, which represents all the limit shapes in a manner similar to how we defined the limit shape in the intra-patient setting (Figure \ref{CCFMN} (right)). 
We optimize the global latent basis (GLB) $\mathcal{Z}=\{Z_{id}\}_{id=1}^P$, representing a functional map from the global shape to each patient's limit shape $L_{id}$, in a manner similar to that used in CCLB, by solving the optimization problem: 
\begin{equation}
    \min_Z \sum_{i,j}\norm{C_{L_i, L_j}Z_i - Z_j } \text{, s.t. } \sum_i Z_i Z_i^T = I.
\end{equation}

Our approach shares the advantage of the CCLB while avoiding the computation of maps between each shape of each collection, which would be computationally expensive. 
Based on the pointwise conversion method proposed by R. Huang et. al. \cite{huang_consistent_2020}, we propose a new pointwise map conversion procedure that utilizes both the CCLB $\{Y_i\}_{i=1}^n$ and the GLB $\{Z_{id}\}_{id=1}^P$ to embed all patient shapes into the global shape spectral domain $\mathcal{Z}$. 
Let $S_i$ be a shape of the patient $id$. We embed the shape $S_i$ into the global shape embedding $\mathcal{Z}$ by using the operator $\zeta_i = \Phi_i Y_i Z_{id}$.
$\zeta_i$ is the composition of three spectral embedding operations: $ \Phi_i$ converts the euclidean representation of the shape $S_i$ into its spectral form in $L^2(S_i)$; $Y_i$ projects the spectral form to the limit shape $L_{id}$ of the patient $P^{id}$; and $Z_{id}$ maps the limit shape coefficients into global shape coefficients.
The point-wise map $T_{ij}$ between two shapes $S_i$ and $S_j$ is obtained by comparing the global shape coefficients of each shape and  solving the following equation:
\begin{equation}
    T_{ij} (p) = \argmin_{q\in S_j} \norm{\zeta_j(q)^T - \zeta_i(p)^T}_2 , ~ \forall p \in S_i.
    \label{pointwisefromcclb_ours}
\end{equation}
In practice, this is done by a nearest neighbors search of each row of $\zeta_i = \Phi_i Y_i Z_{id_i}$ among the rows of $\zeta_j = \Phi_j Y_j Z_{id_j}$.

\subsection{Analysis of shape changes}
\label{deformation_modeling}
We analyze the shape collection data using two complementary approaches. The first is based on the \textit{Characteristic Shape Differences } \cite{huang_limit_2019}, which emphasizes the intrinsic variability within the shape collection.
The second approach makes use of the correspondence  computed between the shapes to quantitatively assess shape changes over the course of radiotherapy sessions.
While the former approach allows the study of deformations
without requiring alignment, it can not directly be used to derive quantitative deformation measurements.
In contrast, the latter, although sensitive to alignment errors, provides displacement vectors that can be used to derive volume change. We present these approaches in detail in the following sections.

\subsubsection{Intrinsic variability analysis}
\label{intrinsicvar}
We conduct an intra- and inter-patient intrinsic shape analysis using the limit and global shape functional representations, respectively constructed in Sections \ref{intra-patient-corr} and \ref{inter-patient-corr}. Specifically, we identify area and conformal deformations on a given shape $S_n$ by employing Characteristic Shape Differences (CSDs) as described by R. Huang et. al. \cite{huang_limit_2019}. They are defined as $D_n^A = Y_n^T Y_n$ for area deformations and $D_n^C = \Lambda_0^+ Y_n^T \Lambda_n Y_n$ for conformal deformations. Here, $\Lambda_0$ and $\Lambda_n$ are the eigenvalues of the limit shape and the shape $S_n$, respectively, and $^+$ denotes the Moore-Penrose pseudo-inverse.
CSDs have many appealing properties, such as their compactness and flexibility, as they rely solely on functional maps and are independent of the shapes' discretization. Another interesting property is their functoriality, which enables the computation of a shape difference $D_{nm}$ between two shapes $S_n$ and $S_m$ as $D_{nm} = Y_n D_n^{-1}D_mY_n^{-1}$. Additionally,
their algebraic nature allows them to be represented as small matrices, thus making them easy to manipulate and compose.
Global variability is modeled in such a way that, after suppressing this variability, the shapes within the collection are aligned as closely as possible. They have shown that the function $\alpha^*$, 
which maximizes the differences in norms across the collection, can be determined as the eigenfunction associated with the largest eigenvalue of the matrix
$E = \sum_{n,m} (D_n-D_m)^2$, where $D_n$ and $D_m$ represent shape differences.
The eigenfunctions $\{\mu_k\}$ of $E$ account for the functional deformations, ordered from the most to the least important. 

To compare shapes from different patient collections in the setting of inter-patient analysis, we first represent them in the common space $\mathcal{Z}$. The CSD $D_{id,n}$ of each shape $S_{id,n}$ from each patient collection $L_{id}$ ($id$=1...$P$, $n$=1...$N$) is transferred to the global shape functional domain $\mathcal{Z}$, to obtain a global CSD $D_{id,n}^g$=$Z_{id}^{-1}D_{id, n}$.
We then compute the global CSD in a way that captures session-to-session variability while minimizing inter-patient differences. 
The global shape variability in our shape collection is computed in the global function domain, with the global CSD and the spectrum of $F = \sum_{i,j} (D_i^g-D_j^g)^2$. 
In addition, we derive the global cross-session variability by maximizing the total shape variability between all shape pairs \{($S_{i,n}$, $S_{j,n}$)\} from different patient collections ($i \neq j$) but belonging to the same session collection $n$, and minimizing the shape variability among all shape pairs $\{(S_{id,n},S_{id,m})\}$, ($n \neq m$) within the same patient collection $id$.

This is implemented by computing the eigenvalues and eigenvectors of:
\begin{equation}
H= \sum_{i,j} \sum_{n} (D_{i,n}^g-D_{j,n}^g)^2 - \sum_{id}\sum_{n,m} (D_{id,n}^g-D_{id,m}^g)^2
\label{eq:H}
\end{equation}

This ensures a robust distinction between inter-session and inter-patient variability, highlighting the regions with the most change across different sessions, while minimizing inter-patient differences.
To obtain a compact visualization that summarizes the primary changes over the mesh, we define a distinctive function in the global shape space,  similar to the method outlined in R. Huang et. al. \cite{huang_limit_2019}. This function is expressed as:
\begin{equation}
    D^{global} = \sum_{k=1}^m \lambda_k^H \mu_k^H ,
    \label{eq:distinctive}
\end{equation}
where $\lambda_k^H$ and $\mu_k^H$ denote the eigenvalues and eigenvectors of $H$.
In Section \ref{results}, results are shown using this visualization in pairwise and groupwise settings. 

\subsubsection{Extrinsic analysis}
\label{sec:extrinsic}

\begin{figure}[htpb]
    \centering
    \includegraphics[width=\textwidth]{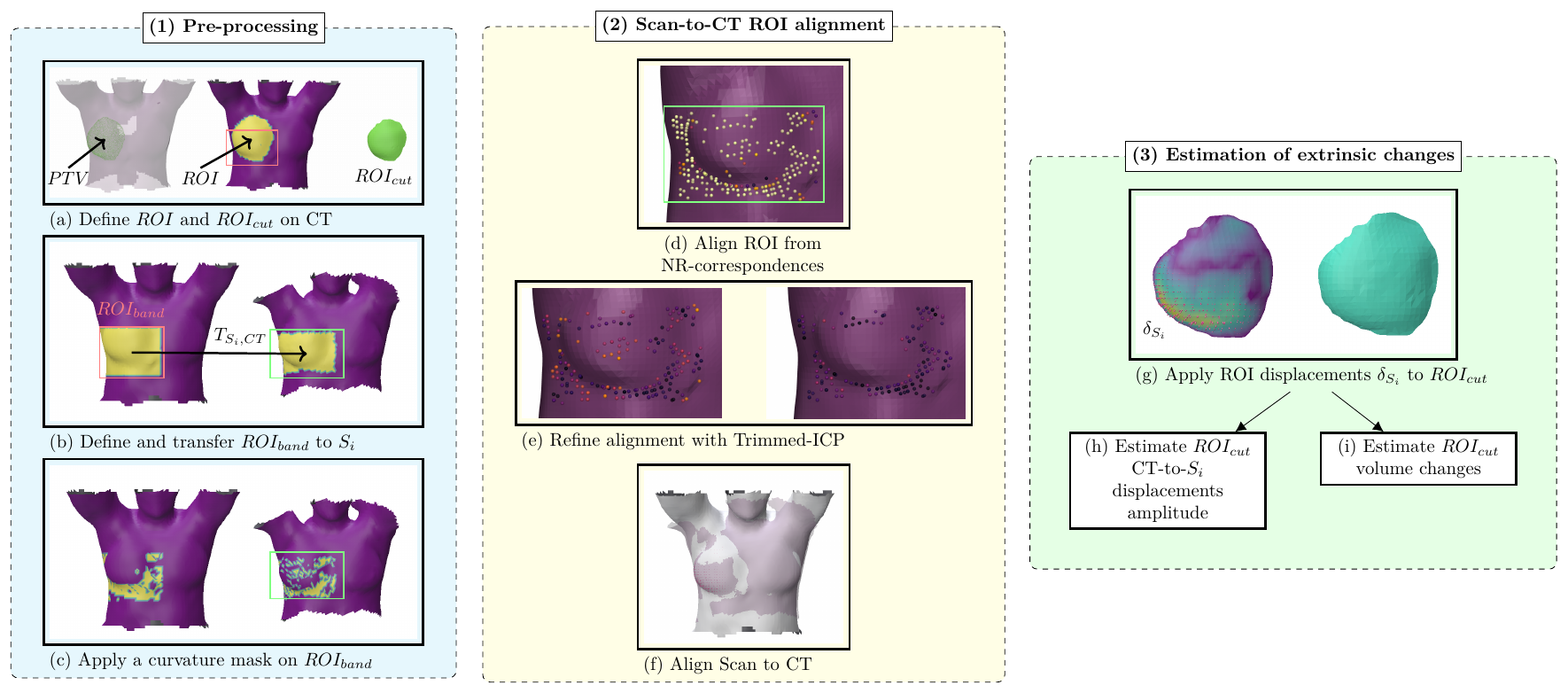} 
    \caption{
Estimation of extrinsic changes. Our approach begins by pre-processing the CT data to isolate the treated breast region (ROI). We then create a derived region, referred to as $ROI_{band}$, to facilitate rigid alignment, and convert the ROI into a closed mesh, termed $ROI_{cut}$ (1). The rigid alignment process (2) involves initializing with non-rigid correspondences, followed by refinement using a trimmed-ICP on the selected points. Finally, we measure the point-to-surface displacement vectors $\delta_{S_i}$ from the CT scans and estimate the resulting volume changes on the $ROI_{cut}$ mesh (3).}
    \label{vol_changes}
\end{figure}

Here we aim to analyze extrinsic changes in the treated bust region (ROI) by comparing the CT skin mesh with  each surface scan in Euclidean space.
We rigidly align the CT skin mesh with each surface scan  using the previously computed non-rigid correspondences. From this alignment, we estimate the per-vertex displacements in the breast region (ROI) of the CT skin mesh, and derive the resulting changes in breast volume.
Figure \ref{vol_changes} illustrates the workflow for estimating extrinsic surface changes in the region of interest, as described below.

\vspace{0.25cm}
\noindent\textbf{(1) Pre-processing.}
We begin by identifying the treated breast region ($ROI$) on the CT skin mesh. The planning target volume ($PTV$) is projected onto the CT skin mesh  using a Euclidean nearest neighbors approach. Each ROI point on the $PTV$ is paired with its closest point on the CT skin mesh, marking the corresponding points as part of the $ROI$. The projected ROI region is then segmented and closed to form a closed mesh, $ROI_{cut}$, representing the treated breast (Figure \ref{vol_changes}(a)).
Next, to maximize the use of the pose- and respiration-invariant part of the mesh in guiding the alignment process, we define a region called $ROI_{band}$ on the CT skin mesh and transfer it to each scan $S_i$ (Figure \ref{vol_changes}(b)). $ROI_{band}$ is defined by the portion of the mesh located within a bounding box determined by the extreme points of the $ROI$. $ROI_{band}$ includes regions such as the rib cage and sternum, which are beneficial for rigid alignment. The upper boundary of this region is lowered to reduce the inclusion of points from the armpit area, which is more prone to non-rigid deformations. Finally, we compute $ROI_{mask} = ROI_{band} - ROI$ to retain only the points outside the breast region, which is likely to undergo non-rigid deformations across sessions (Figure \ref{vol_changes}(c)).

\vspace{0.25cm}
\noindent\textbf{(2) Rigid alignment from non-rigid correspondences.}
We then align the CT skin mesh with each scan mesh $S_i$ using a goal function that minimizes the sum of distances between the two $ROI_{mask}$s (Figure \ref{vol_changes}(d)).
The rigid alignment thus obtained is further refined using a point-to-surface Trimmed Iterative Closest Point (ICP) algorithm (Figure \ref{vol_changes}(e)). This step improves the alignment by iteratively excluding the farthest scan points relative to the CT, effectively removing points that could contribute to non-rigid deformations.

\vspace{0.25cm}
\noindent\textbf{(3) Estimation of breast changes.}
For each optical scan $S_i$, we define a vector field $\delta_{S_i}$ on the CT skin breast surface by connecting each vertex $p$ on the CT skin to its closest point $q$ on the $S_i$. Such vertex-to-surface approach allows the displacements to be robust to the discretization of the aligned scan.

We then estimate the volume changes by deforming the closed mesh $ROI_{cut}$ onto $ROI_{S_i} = ROI_{cut} + \delta_{S_i}$, where $\delta_{S_i}$ is the vector field calculated for each clinical trial session $S_i$. From this deformation, we calculate the relative volume differences $\Delta V^{rel}(S_i)$ between session $S_i$ and the initial volume computed on $ROI_{cut}$ using the following formula:
\begin{equation}
    \Delta V^{rel}(S_i) = 
    \frac{V(ROI_{S_i}) - V(ROI_{cut})}{V(ROI_{cut})}.
\end{equation}

\section{Implementation}
\noindent\textbf{Shapes pre-processing.}
We first remesh the surface scan meshes to approximately 10K vertices using the Meshlab \cite{cignoni_meshlab_2008} VCG (Visual Computing Group) surface reconstruction filter. In addition, we make sure the busts are composed of only one connected component by using Meshlab's ``Cleaning and Repairing" filters. We also extract from the planning CT scan the point cloud representing the patient's torso surface and remesh it to obtain a 10K vertice mesh that we denote as \textit{CT skin mesh}.
This preprocessing generates for each patient a dozen of torso meshes with similar connectivity that we use as input to study the intra-patient deformation.

\vspace{0.25cm}
\noindent\textbf{Shape matching parameters.}
In our experiments, we calculated initial pairwise functional maps with a size of \( k=65 \), used limit shapes with a size of \( k_l=52 \), and finally, employed a global shape size of \( m=20 \). These parameters were selected based on preliminary tests to ensure that the spectral embeddings were large enough to effectively represent the shapes, while also balancing the need to avoid excessive computational costs and potential inaccuracies caused by high-frequency noise in the optimization process.

\vspace{0.25cm}
\noindent\textbf{Shape alignments.}
For the rigid alignment, we utilize the Procrustes and Iterative Closest Point (ICP) algorithms provided by the ``trimesh" Python library. The trimesh implementation of ICP allows us to use a mesh as the target and to measure vertex-to-surface distances rather than vertex-to-vertex distances. In practice, the $ROI_{band}$ region is defined by shifting the top of the ROI 3 cm down along the z-coordinate to exclude part of the armpit region, and shifting the bottom 6 cm down along the z-coordinate to include the ribs region. In addition, 3 cm are added to both sides along the x-coordinate to ensure consistent treatment for patients with either right or left breast cancer. Finally, a 5 cm shift along the y-coordinate (depth) is applied to include points from the ribs on the treated breast side. 
We ran 100 iterations of ICP, applying trimming every 5 iterations to remove the two percent of points with the highest point-to-surface distances.

\section{Results}
\label{results}
We present our results in two parts. First, we conduct an intrinsic shape analysis to qualitatively highlight changes in the bust during treatment.  Next, we  utilize non-rigid correspondences to compute displacement vectors from the CT skin to the multiple shapes captured throughout the treatment, which are then used to quantify the evolution of breast.

\subsection{Intrinsic variability analysis}
We utilize the global shape functional domain and the related global CSD presented in section \ref{intrinsicvar} to explore the shape variability between the woman busts at three different levels: pairwise, 
intra-patient, and inter-patient.

\begin{figure}[htpb]
    \centering
    \includegraphics[width=0.75\textwidth]{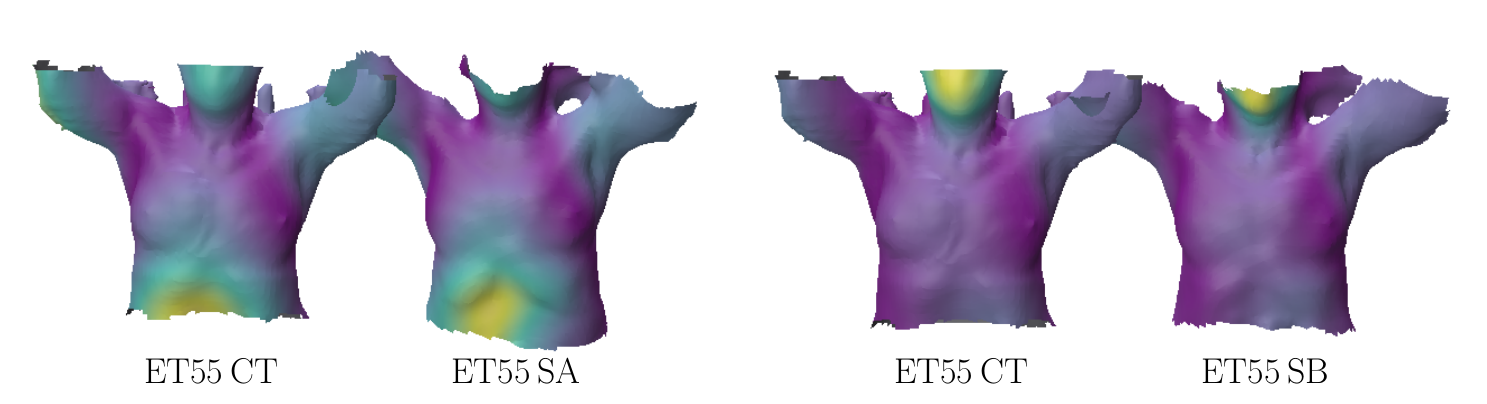} 
    \caption{Pairwise conformal distinctive functions computed in the global shape space $G$. The A and B sessions are compared to the centering CT acquisition, highlighting in yellow the most distinctive region for conformal distortion.}
\label{fig:conformal_pairwise}
\end{figure}

\noindent\textbf{Pairwise  analysis.}
Figure \ref{fig:conformal_pairwise} shows an example of the pairwise analysis by using the global shape space $G$. The two surface scans (A, B) of a subject are compared with the CT skin surface acquired on the same day.
Conformal distinctive functions are computed using Equation (\ref{eq:distinctive}) and visualized as colormaps on two shape pairs.
We observe that even when the acquisitions are performed on the same day, the bust undergoes deformations, such as the belly movement due to respiratory motion (Figure \ref{fig:conformal_pairwise}(left)) or head movement (Figure \ref{fig:conformal_pairwise}(right)). 

\begin{figure}[htpb]
    \centering
    \includegraphics[width=0.9\textwidth]{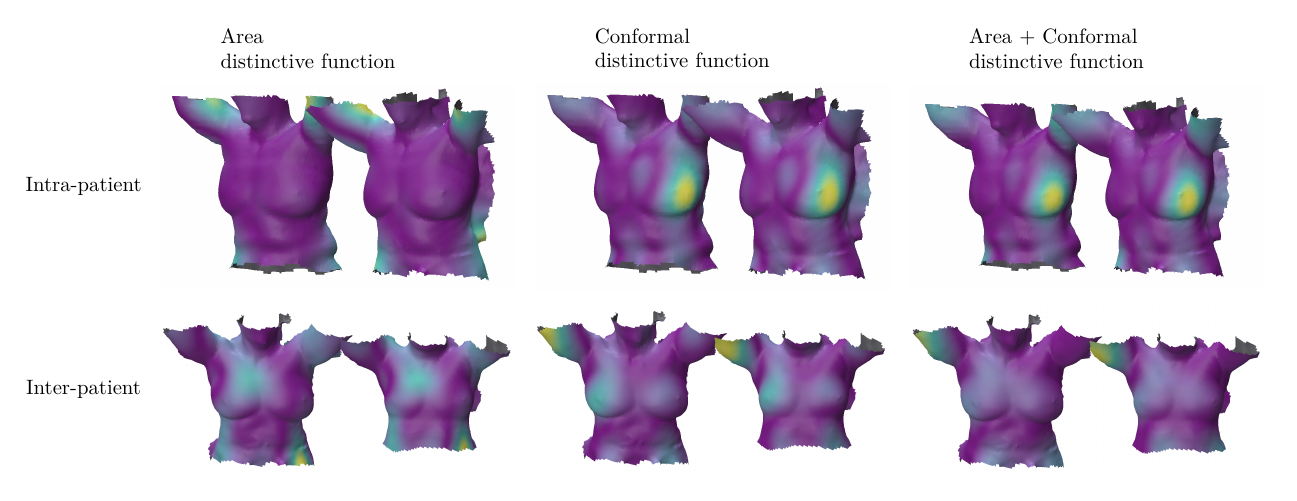} 
    \caption{Pairwise distinctive functions computed in the global shape space $G$, and truncated at the last index are shown in different configurations. The first line shows distinctive functions between acquisitions CT and $S_{25}$ of the same patient (BF37). The second row compares the sessions $S_{25}$ of two different patients (BF37 and PW44). Regions with the most distortion are highlighted in yellow. }
    \label{distinctive_pairwise_table}
\end{figure}

Figure \ref{distinctive_pairwise_table} illustrates intrinsic analyses that can be performed on two shapes.
Users can compare a patient's CT scan with the surface scan from a specific session (first row), or shapes of two different patients (second row). 
Each column highlights regions of major distortions (yellow) for a given metric: area-based (first column), conformal-based (second column), and combined area- and conformal-based (third column).

\begin{figure}[htpb]
    \centering
    \includegraphics[width=0.9\textwidth]{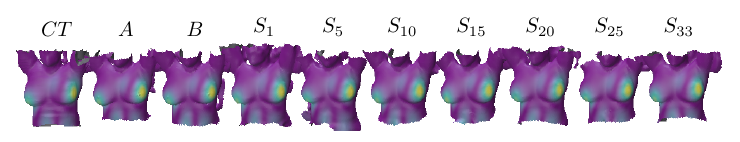} 
    \caption{Intra-patient intrinsic analysis (patient SB50). The 2nd conformal global variability function is shown on each shape of the patient, showing conformal distortion in the treated (left) breast region.}
    \label{intra-patient-SB50}
\end{figure}

\begin{figure}[htpb]
    \centering
    \includegraphics[width=1.0\textwidth]{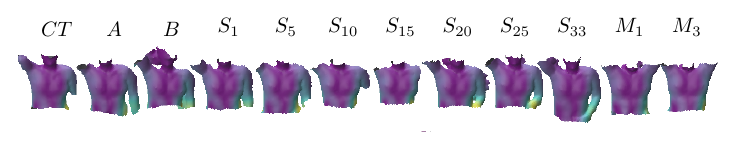} 
    \caption{Intra-patient intrinsic analysis (patient SD19). The 19th conformal global variability function is shown on each shape of the patient, showing conformal distortion in the left arm of the patient.}
    \label{intra-patient-SD19}
\end{figure}

\noindent\textbf{Intra-patient analysis.}
We also can compare the shapes in an intra-patient manner, always using the distinctive functions or with the function $\alpha^*$ maximizing the changes of norms.
Figures \ref{intra-patient-SB50} and \ref{intra-patient-SD19} respectively show the 3rd and 19th eigenvectors of the patient collection's conformal CSD, out of a total of 20. On SB50, both breast regions are highlighted, with a higher variability observed in the treated (left) breast. The intrinsic analysis captured the conformal variability in the breast regions, when considering all the shapes in the collection. However as 2nd out of 20, this variability is not the most prominent one, and regions like the arms, the belly and noisy are shown with a higher conformal variability.
SD19 highest variability is observed in the region of the left arm of the patient. 
We can observe that this patient likely experienced an issue with arm mobility during the treatment, which was addressed one and three months after the radiotherapy.

\begin{figure}[htpb]
    \centering
    \includegraphics[width=1.0\textwidth]{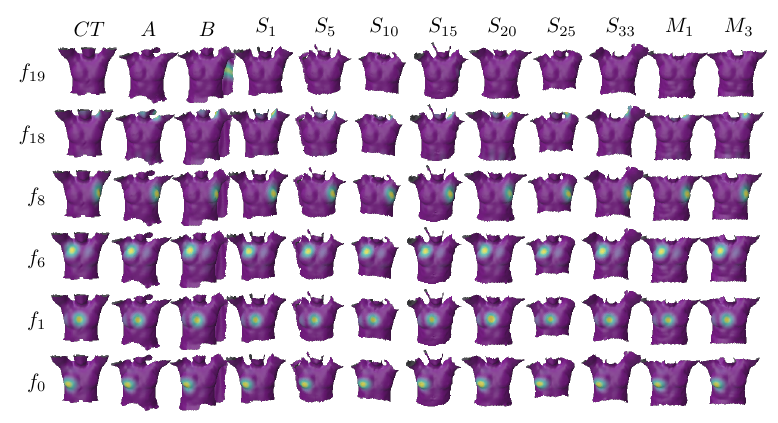} 
    \caption{Distinctive regions $f_i$ of patient LB06. Each line shows a region $f_i$ where there is a global area variability when taking into account all sessions.}
    \label{area_global_var}
\end{figure}

In general, we observed in our experiments that the most distinctive regions, so the eigenvectors associated with the highest eigenvalues were located in regions where the coverage is not the same for all the shapes in the collection as depicted in Figure \ref{area_global_var}.

The global area variability is shown on the whole bust with respect to all other acquisitions of one patient. Each row represents a distinct region $f_i$ where area variability has been observed, ranging from the most significant $f_{19}$ to the least significant $f_0$, in terms of variability across all sessions.
We can observe that while the most distinctive regions $f_{19}$ and $f_{18}$ highlight the table and the shoulders that are not common to all shapes, the latter ones gives interesting information. Indeed, $f_8$ depicts area changes in the treated breast, $f_6$ and $f_0$ changes in the other breast, and  $f_1$ on the sternum.
For this reason, in case of noisy data, $\alpha^*$ is not the most interesting region to highlight. Because of this observation, we looked at the various distinctive functions $f_i$ and only kept those not highlighting noisy regions like the table as shown in Figure \ref{area_global_var} with the distinctive function $f_{19}$ detecting shape variability with respect to the table noise on acquisition $B$.
In practice, we truncated the distinctive functions at the first index containing noise, eliminating the noise variability from our bust change analysis.

\begin{figure*}[htpb]
    \centering
    \includegraphics[width=\textwidth]{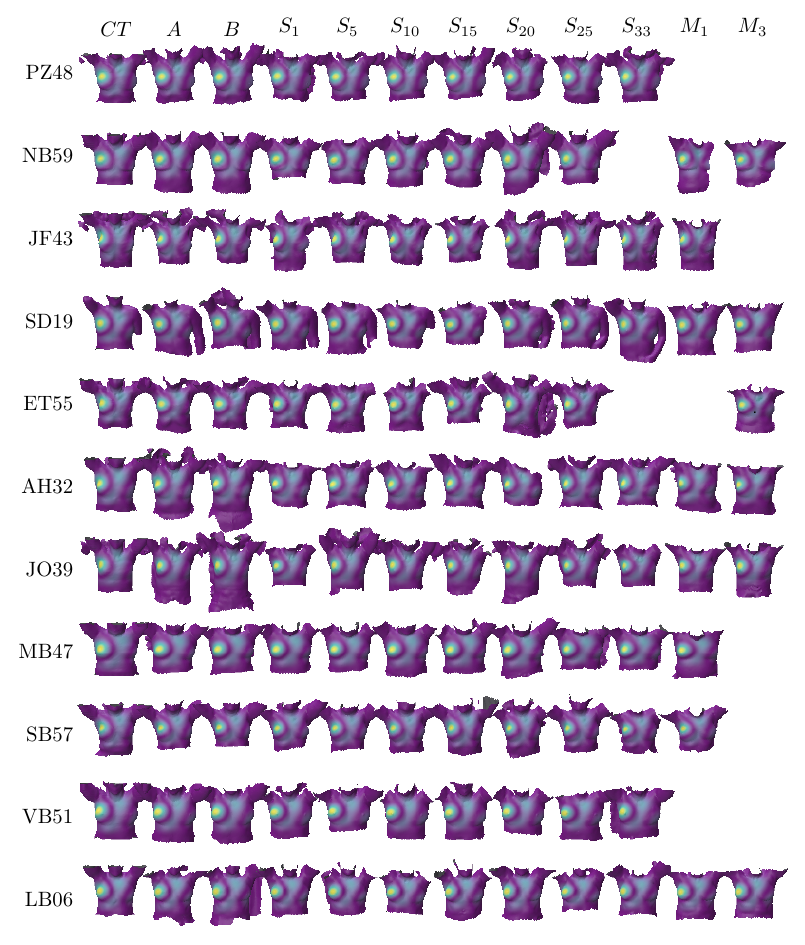} 
    \caption{Inter-patient global area variability. The global area distinctive function, truncated at the 9th index over 20 is shown in a color map (yellow is the highest) on all the shapes.}
    \label{inter-patient-right-breast}
\end{figure*}

\begin{figure*}[htpb]
    \centering
    \includegraphics[width=\textwidth]{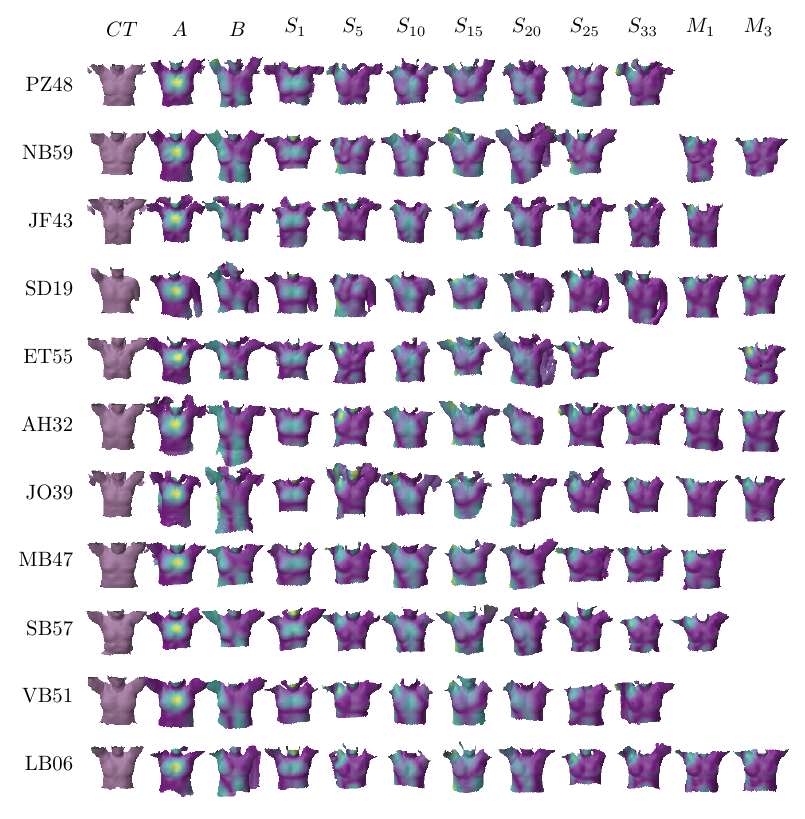} 
    \caption{Inter-session combined (area and conformal) variability on patient treated for right-breast radiotherapy. The distinctive function, truncated at the 17th index over 20 is shown on all the shapes, with higher values in yellow.}
    \label{inter-patient-right-breast-inter-sessions}
\end{figure*}

\begin{figure*}[htpb]
    \centering
    \includegraphics[width=0.97\textwidth]{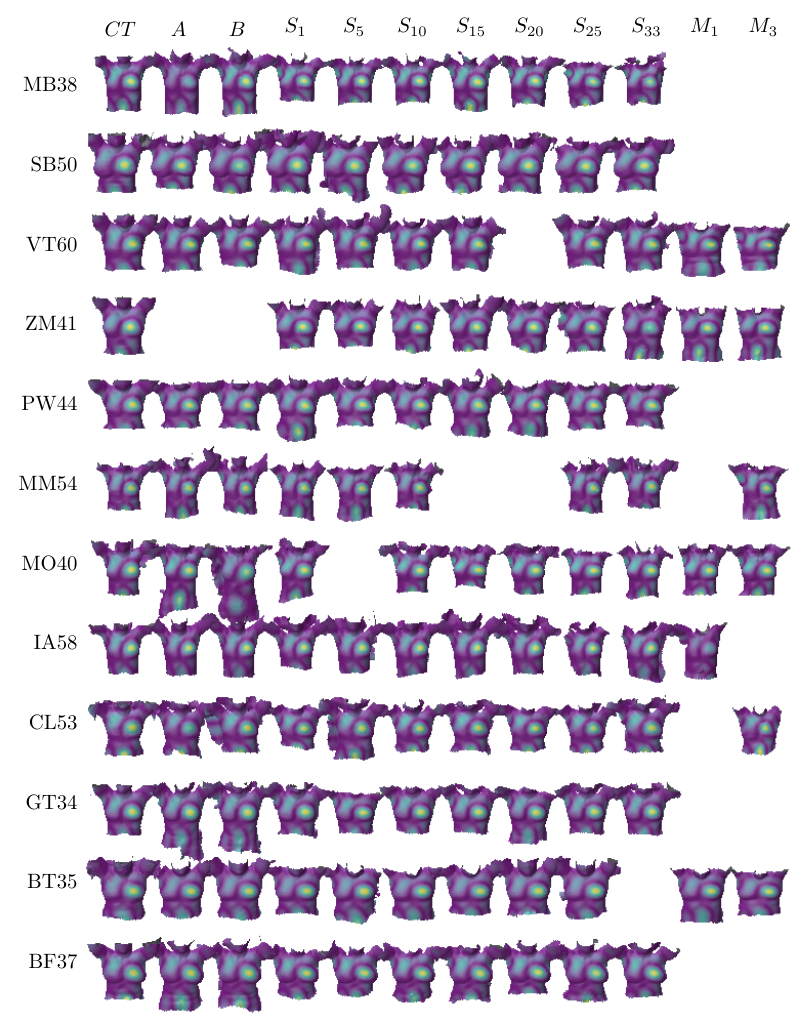} 
    \caption{Inter-patient cross-session conformal variability in patients undergoing treatment for the left breast. The cross-session conformal distinctive function, truncated at the 13th index over 20, is shown on all the shapes with highly distinctive regions in yellow.}
    \label{inter-patient-left-breast}
\end{figure*}

\vspace{0.25cm}
\noindent\textbf{Inter-patient analysis.}
Figure \ref{inter-patient-right-breast} shows the inter-patient shape variability computed on the shape collection of the right breast from 11 patients undergoing radiotherapy.
The intrinsic variability captures area changes in the treated breast region, indicating that the treated breast tends to vary during the course of the treatment. However, this global analysis also considered inter-patient area variability, as it incorporated all the shapes of each patient into a single shape collection.

Our framework allows to personalize the shape differences the user wants to study. Shapes can be labeled as belonging to specific collections that can then be compared with equation (\ref{eq:H}).
To analyze our radiotherapy treatment acquisitions, we conducted a cross-collection analysis across different sessions to identify shape changes between sessions, while minimizing inter-patient variability.
The distinctive functions (Equation \ref{eq:distinctive}) was used to highlight regions that contribute to the shape variability, as shown in Figure \ref{inter-patient-right-breast-inter-sessions}. For session A, we observe that the sternum region is highlighted, indicating variability likely caused by the patient's respiratory motion. For later sessions, like $S_{25}$ to $M_3$, we observe the main variability is located between the treated breast and the armpit on the same side. This behavior highlights potential arm movements of the patients during these irradiation sessions.

Figure \ref{inter-patient-left-breast} illustrates an inter-collection shape analysis, where each collection comprises session acquisitions from all patients. This way, we have 12 shape collections, one for each session, and apply equation (\ref{eq:H}) to measure the variability between the sessions, while minimizing the variations between the patients within the session collection. This analysis highlighted the treated breast region, as the region undergoing the most conformal distortions during the treatment.

The results presented above demonstrated that intrinsic analysis can be a great tool to study variations in shape collections, without requiring alignment. In addition, the functional map construction used for this analysis is highly flexible and scalable, allowing new patient shapes and collections to be incrementally added without the need to reprocess existing data.
However, by its intrinsic nature, such analysis is not suited to define vector field deformations, that can be required in some applications.

\subsection{Extrinsic analysis}

\begin{figure*}[htpb]
    \centering
    \includegraphics[width=\textwidth]{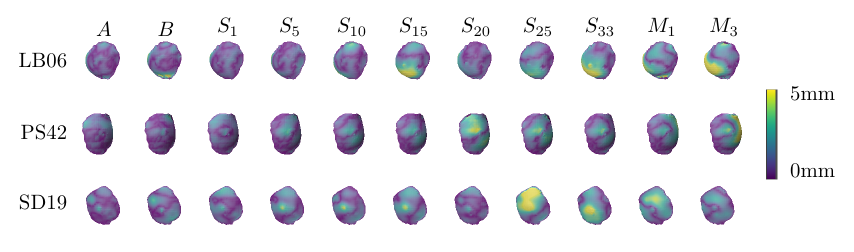} 
    \caption{$ROI_{cut}$ has been deformed to match the $ROI_{S_i}$. The colormap visualizes the amplitude of the displacement vectors.}
    \label{extrinsic-deltas}
\end{figure*}

We conducted an extrinsic shape analysis, as described in Section \ref{sec:extrinsic}, to evaluate surface-level shape changes in the treated breast region.
We began by applying the rigid alignment method described in Section \ref{sec:extrinsic} to all available patient data. 
We then measured the length of the deformation vectors from the CT to the scan within the region of interest. 
Figure \ref{extrinsic-deltas} presents the extracted breast region, denoted as $ROI_{cut}$, deformed to match the various sessions on three different patients. Yellow regions indicate areas of higher displacement amplitudes.
The deformations vary depending on the session and patient; for example, LB06 (first row) exhibits localized deformations at the bottom of the breast in sessions $S_{15}$, $S_{25}$, and $S_{33}$, while PS42 (second row) shows more widespread variations across the entire breast. For patient SD19 (third row), the presence of nipples is depicted as a yellow point in the middle of the deformed breasts, a result of the absence of the nipple in the pre-processed centering CT skin mesh.
This qualitative analysis effectively highlights the regions where the treated breast experiences the most deformation, visually showing an increase in breast displacements for SD19 during session $S_{25}$.

To quantify the deformations in the breast region defined in Section \ref{intra-patient-corr}, we calculated the average displacement vector amplitudes, as shown in Table \ref{tab:deltas_amplitude}.
The average and standard deviation of displacement lengths are presented for each (patient, session) pair in millimeters. 
Notably, session $S_{25}$ for patient SD19 exhibits the highest deformation, with average displacements around 3 mm in the breast region, compared to approximately 1 mm in other sessions. This quantitative analysis confirms the increased deformations observed qualitatively for SD19 during this session.
Overall, we find that the treated breast undergoes non-rigid surface displacements of approximately 2 mm on average, with increased variability observed at the end of the treatment (session $S_{25}$) and in the months following the treatment (M1 and M3).

\vspace{0.35cm}
\begin{table}[htpb]
    \centering
    \resizebox{\textwidth}{!}{%
\begin{tabular}{|l|l|lllllllllll|l|}
\hline
 & CT ROI vol (mL) & SA & SB & S1 & S5 & S10 & S15 & S20 & S25 & S33 & M1 & M3 & Sessions avg \\
 \hline
SD19 & 301.71 & 0.64±0.58 & 1.01±0.77 & 0.95±0.74 & 1.14±0.83 & 1.03±0.84 & 1.37±0.92 & 0.70±0.55 & \underline{\textbf{3.06±1.90}} & \textbf{2.19±1.37} & \underline{1.95±1.16} & 1.18±0.71 & 1.46±0.73 \\
JO39 & 333.40 & 1.48±0.96 & 0.90±0.59 & 1.55±1.01 & 1.37±0.99 & \underline{1.99±1.40} & 1.50±0.95 & 1.83±1.40 & \textbf{3.07±2.37} & 1.57±1.37 & 1.48±1.01 & \underline{\textbf{3.62±2.74}} & 1.89±0.83 \\
MB52 & 369.17 & 0.83±0.74 & 0.65±0.57 & 1.26±0.98 & 1.22±1.20 & 0.81±0.72 & 0.89±0.68 & 0.73±0.57 & \underline{1.32±1.34} & \textbf{1.50±1.21} & 1.24±0.85 & \underline{\textbf{2.74±2.55}} & 1.24±0.60 \\
NB59 & 380.83 & 0.91±0.79 & 1.14±1.00 & \underline{1.88±1.30} & 1.03±0.77 & 1.12±0.70 & 0.98±0.81 & 1.09±0.96 & 1.47±1.12 & - & \textbf{1.98±1.54} & \underline{\textbf{2.00±1.50}} & 1.41±0.43 \\
PS42 & 397.75 & 0.80±0.59 & 0.56±0.66 & 0.76±0.50 & 0.81±0.72 & 0.84±0.68 & 0.88±0.72 & \textbf{2.08±1.24} & 1.44±1.07 & 0.91±0.78 & \underline{1.47±1.20} & \underline{\textbf{2.26±1.93}} & 1.20±0.59 \\
MO40 & 435.22 & 0.96±0.69 & \underline{\textbf{3.18±1.92}} & 1.06±0.64 & - & 0.65±0.51 & 0.86±0.92 & 1.11±0.96 & 0.83±0.86 & 0.81±0.63 & \underline{1.75±1.22} & \textbf{2.00±1.51} & 1.36±0.82 \\
MB38 & 482.86 & 0.45±0.32 & 0.44±0.35 & 1.74±1.38 & 2.48±2.48 & 2.34±1.33 & 1.10±1.00 & \textbf{4.15±3.18} & \underline{\textbf{5.58±4.65}} & \underline{3.71±2.67} & - & - & \underline{\textbf{2.69±1.70}} \\
LB06 & 489.94 & 0.98±1.22 & 0.95±1.01 & 0.69±0.70 & 0.94±1.09 & 1.29±1.34 & \underline{1.99±1.40} & 0.93±0.95 & 1.78±1.30 & 1.61±1.17 & \textbf{2.25±1.83} & \underline{\textbf{2.93±2.58}} & 1.54±0.71 \\
PZ48 & 495.18 & 0.81±0.67 & \textbf{0.94±0.75} & 0.59±0.55 & 0.80±0.66 & 0.76±0.62 & \underline{\textbf{1.19±0.94}} & \underline{0.82±0.71} & - & 0.58±0.54 & - & - & 0.81±0.21 \\
IA58 & 497.86 & 1.22±0.73 & 0.92±0.66 & \textbf{2.37±1.79} & 1.88±1.28 & 1.23±0.89 & 0.93±0.73 & 1.54±0.95 & \underline{2.06±1.40} & 1.10±0.81 & \underline{\textbf{2.80±2.24}} & - & 1.65±0.67 \\
PW44 & 513.38 & 1.52±0.94 & 0.73±0.58 & 1.11±1.15 & 1.52±1.07 & 1.19±1.08 & 0.89±0.73 & \underline{2.03±1.70} & \textbf{2.05±1.74} & \underline{\textbf{2.20±1.43}} & - & - & 1.46±0.57 \\
GT34 & 555.85 & 1.04±0.75 & 1.03±0.70 & 0.93±0.64 & 0.95±0.63 & \textbf{1.52±1.15} & 1.09±0.80 & \underline{\textbf{1.92±1.31}} & \underline{1.40±1.15} & 1.14±0.94 & - & - & 1.25±0.34 \\
JF43 & 575.26 & \textbf{2.11±1.85} & 1.48±1.07 & \underline{\textbf{2.77±2.50}} & 0.96±0.87 & \underline{1.71±1.38} & 1.57±1.52 & 1.66±1.82 & 1.68±1.36 & 1.08±0.83 & 1.55±1.44 & - & 1.61±0.51 \\
ZM41 & 686.05 & - & - & \textbf{1.09±0.89} & 0.86±0.72 & 0.91±0.87 & 0.98±0.98 & 0.90±0.67 & 0.78±0.69 & 0.96±0.69 & \underline{\textbf{2.88±2.80}} & \underline{1.01±0.99} & 1.15±0.65 \\
MM54 & 698.51 & 0.65±0.58 & 0.98±0.90 & 1.46±1.77 & 1.02±1.05 & \underline{\textbf{2.25±1.76}} & - & - & \textbf{1.94±1.84} & \underline{1.77±1.35} & - & 1.76±1.90 & 1.60±0.47 \\
SD56 & 700.37 & 1.62±1.40 & 1.37±1.21 & \textbf{2.13±1.92} & \underline{1.85±1.23} & 1.24±1.01 & 0.94±0.97 & 1.46±1.49 & 1.64±1.44 & - & - & \underline{\textbf{4.60±4.48}} & 1.90±1.15 \\
BT35 & 722.53 & 1.10±1.05 & 1.10±1.03 & 1.35±1.29 & 1.49±1.53 & \textbf{2.23±1.99} & 2.07±1.95 & 1.93±1.98 & \underline{2.14±2.29} & - & 1.42±1.44 & \underline{\textbf{3.38±3.15}} & 1.90±0.68 \\
MB47 & 757.91 & 1.97±1.66 & 1.54±1.20 & \textbf{2.89±1.85} & 1.99±1.55 & 2.50±2.17 & 1.69±1.21 & \underline{2.53±2.17} & 1.81±1.46 & 2.18±1.42 & \underline{\textbf{3.00±2.06}} & - & \textbf{2.24±0.52} \\
LD49 & 808.94 & 0.60±0.53 & 0.51±0.43 & 1.30±1.06 & \textbf{1.66±1.47} & 0.93±0.98 & \underline{1.53±1.04} & 1.28±1.10 & 0.99±0.88 & - & 1.35±1.13 & \underline{\textbf{6.82±6.04}} & 1.82±1.91 \\
AF45 & 877.28 & \underline{2.21±2.02} & \textbf{2.26±1.89} & 1.51±1.27 & 1.63±1.14 & 1.68±1.50 & 1.33±1.16 & 1.56±1.38 & 1.27±1.05 & \underline{\textbf{2.62±1.82}} & - & - & 1.73±0.47 \\
SD03 & 1185.07 & 1.14±0.88 & 1.10±1.12 & 2.23±1.70 & 1.60±0.98 & \underline{2.39±1.43} & 0.75±0.70 & \underline{\textbf{3.23±2.99}} & 1.14±1.10 & - & \textbf{2.59±1.84} & 2.35±1.85 & \underline{1.93±0.83} \\
SB50 & 1210.52 & \textbf{2.47±1.86} & \underline{\textbf{2.52±1.68}} & 1.59±1.20 & 1.31±0.96 & 1.78±1.39 & \underline{1.81±1.24} & 1.79±1.38 & 1.23±1.04 & 1.69±1.23 & - & - & 1.71±0.39 \\
\hline
Patients avg & 612.53 & 1.21±0.58 & 1.20±0.69 & 1.51±0.64 & 1.36±0.45 & 1.47±0.60 & 1.25±0.40 & 1.68±0.85 & \underline{1.84±1.05} & 1.62±0.78 & \textbf{1.98±0.62} & \underline{\textbf{2.82±1.56}} & 1.62±0.72 \\
\hline
\end{tabular}
}
    \caption{ROI displacements amplitude in millimeters. The average and standard deviation of the displacement vectors length are provided for each patient session. The three highest values are highlighted as follows: the \underline{\textbf{highest value}} is bold and underlined, the \textbf{second-highest} is bold, and the \underline{third-highest} is underlined.}
    \label{tab:deltas_amplitude}
\end{table}
\vspace{0.35cm}

Table \ref{tab:relative_volume_CT} presents the relative volume changes (\%) with respect to the reference mesh, observed across various sessions. 
Notably, session $S_{25}$ for patient SD19 shows the most significant change, with a potential 18\% increase in breast volume.
It is important to note that extreme values may reflect significant breast changes or potential issues with the alignment process. In either case, this approach offers a useful tool for automatically monitoring extrinsic changes in patients during each session and identifying sessions that deviate from the norm.

\vspace{0.25cm}
\begin{table*}[htpb]
    \centering
    \resizebox{\textwidth}{!}{%
\begin{tabular}{|l|l|lllllllllll|l|}
\hline
 & CT ROI vol (mL) & SA & SB & S1 & S5 & S10 & S15 & S20 & S25 & S33 & M1 & M3 & Sessions avg \\
 \hline
SD19 & 301.71 & 0.98 & 0.52 & 4.00 & 5.52 & 2.02 & 6.45 & 1.87 & \underline{\textbf{18.23}} & \textbf{13.50} & \underline{-7.65} & -4.68 & \underline{6.44±3.14} \\
JO39 & 333.40 & -5.29 & -4.03 & -0.43 & -7.44 & \underline{-10.54} & -4.86 & -9.47 & \textbf{-18.20} & -7.89 & -6.17 & \underline{\textbf{-21.82}} & \underline{\textbf{9.09±2.19}} \\
MB52 & 369.17 & -2.54 & -1.38 & \textbf{-5.61} & 3.70 & -1.90 & 3.34 & -0.05 & \underline{4.16} & 0.26 & 0.98 & \underline{\textbf{15.18}} & 3.66±3.49 \\
NB59 & 380.83 & -0.53 & 2.07 & \textbf{7.70} & 2.98 & 3.33 & 0.87 & 1.72 & \underline{4.37} & - & 2.50 & \underline{\textbf{7.99}} & 3.72±2.07 \\
PS42 & 397.75 & 3.20 & -1.97 & 3.21 & -2.01 & 0.48 & 0.70 & \underline{-3.56} & -0.67 & -0.24 & \textbf{6.02} & \underline{\textbf{9.45}} & 2.83±1.72 \\
MO40 & 435.22 & 0.30 & \textbf{6.76} & -1.95 & - & 0.15 & -1.07 & -1.14 & -1.94 & -0.70 & \underline{-2.73} & \underline{\textbf{7.10}} & 2.62±2.07 \\
MB38 & 482.86 & 1.58 & 1.16 & 1.52 & \underline{8.56} & -4.09 & 1.94 & \textbf{-12.99} & \underline{\textbf{-22.12}} & -4.72 & - & - & \textbf{7.14±1.50} \\
LB06 & 489.94 & 1.65 & -0.28 & 0.92 & 0.97 & 0.29 & \textbf{6.06} & 2.18 & \underline{5.29} & 3.79 & 2.37 & \underline{\textbf{10.07}} & 3.22±1.84 \\
PZ48 & 495.18 & \underline{-1.25} & 0.56 & -0.27 & \underline{\textbf{-2.57}} & \textbf{-1.79} & -0.29 & -1.06 & - & -0.50 & - & - & 1.01±6.48 \\
IA58 & 497.86 & 2.38 & 1.81 & \underline{\textbf{5.02}} & \underline{4.50} & \textbf{4.54} & 0.25 & 3.68 & 0.02 & 3.09 & -2.52 & - & 2.82±6.90 \\
PW44 & 513.38 & -1.40 & 0.39 & \underline{3.79} & -1.76 & -2.21 & 0.98 & \textbf{4.08} & \underline{\textbf{6.30}} & -0.45 & - & - & 2.50±7.27 \\
GT34 & 555.85 & 0.11 & 0.01 & 0.59 & 0.53 & \underline{-3.45} & -1.33 & \underline{\textbf{-5.25}} & \textbf{-4.32} & -0.13 & - & - & 1.95±2.39 \\
JF43 & 575.26 & 2.86 & \textbf{3.76} & \underline{\textbf{6.99}} & 0.50 & 3.53 & 3.13 & -0.04 & \underline{3.62} & 1.81 & -0.63 & - & 2.67±4.43 \\
ZM41 & 686.05 & - & - & 1.28 & -0.70 & 0.70 & \textbf{2.12} & 0.77 & 0.50 & \underline{1.62} & \underline{\textbf{11.42}} & -0.94 & 2.23±2.56 \\
MM54 & 698.51 & -1.38 & -0.43 & 1.74 & \textbf{-3.01} & -1.94 & - & - & \underline{\textbf{-5.79}} & -1.27 & - & \underline{2.79} & 2.42±2.54 \\
SD56 & 700.37 & \underline{6.97} & 5.87 & \textbf{9.28} & 4.47 & 1.57 & -0.61 & 1.21 & 3.72 & - & - & \underline{\textbf{-14.81}} & 5.19±2.94 \\
BT35 & 722.53 & -1.59 & -1.89 & -3.12 & -4.22 & -5.83 & \underline{-6.06} & -4.77 & \textbf{-6.37} & - & -3.97 & \underline{\textbf{-8.94}} & 5.02±2.08 \\
MB47 & 757.91 & -3.32 & -1.25 & -4.21 & 1.09 & \textbf{-6.52} & -1.30 & -3.53 & 2.00 & \underline{-4.37} & \underline{\textbf{-7.65}} & - & 3.55±0.87 \\
LD49 & 808.94 & -1.27 & -0.70 & 2.86 & \textbf{4.63} & 2.37 & 0.49 & \underline{3.52} & 2.37 & - & 2.61 & \underline{\textbf{22.78}} & 4.70±0.57 \\
AF45 & 877.28 & \textbf{-2.83} & \underline{\textbf{-4.97}} & \underline{2.00} & -0.08 & 1.88 & 0.81 & -0.91 & -0.81 & 1.13 & - & - & 1.57±2.14 \\
SD03 & 1185.07 & 2.63 & -2.51 & \underline{\textbf{5.36}} & 0.55 & -0.14 & -0.49 & \underline{-4.38} & -0.25 & - & \textbf{5.07} & -2.59 & 2.37±5.54 \\
SB50 & 1210.52 & -1.59 & -0.05 & -0.90 & -1.00 & \textbf{-1.69} & -1.26 & \underline{\textbf{-1.79}} & -1.15 & \underline{1.64} & - & - & 1.18±4.81 \\
\hline
Patients avg & 612.53 & 2.17±1.62 & 2.02±1.97 & 3.31±2.51 & 2.89±2.37 & 2.77±2.45 & 2.11±2.05 & 3.24±3.13 & \textbf{5.34±6.30} & 2.77±3.46 & \underline{4.45±3.04} & \underline{\textbf{9.93±6.98}} & 3.54±3.16 \\
\hline
\end{tabular}
}
    \caption{Treated Breast ROI relative volume evolution in \%. The volume initial volume $V(ROI_{closed})$ is given for each patient in the first column. The other columns report the relative volume differences $V(ROI_{S_i})$ for each session, relatively to the volume of the CT. The three highest values are highlighted as follows: the \underline{\textbf{highest value}} is bold and underlined, the \textbf{second-highest} is bold, and the \underline{third-highest} is underlined.}
\label{tab:relative_volume_CT}
\end{table*}

\vspace{0.5cm}
\section{Conclusion}
We developed a comprehensive pipeline to explore shape variability in a pairwise, collection, and cross-collection manner for analyzing breast shape changes during post-operative radiotherapy. Based on a functional representation, 
it also introduces Global Latent Bases,
enabling the representation of shapes from different collections within a shared domain, favoring shape matching and analysis. 
This work has been successfully applied to a challenging clinical trial dataset consisting of several hundred shapes of women's torso with varying coverage and noise. Our analysis results demonstrated that the breast region undergoes non-negligible shape and volume changes during the therapy, with average changes of approximately 10\%.

To the best of our knowledge, this is the first trial to utilize functional maps on such a unique dataset. It is also the first computational approach to study the breast shape evolution during the radiotherapy both quantitatively and qualitatively.  
At the same time, it represents the first computational approach to analyze breast shape evolution during radiotherapy from both quantitative and qualitative perspectives.
We believe that the detailed understanding of shape change can contribute to an improve personalized radiotherapy treatment, such dynamic dose planning over the treatment or patient positioning.
Investigating the dosimetric impact of these deformations would be of significant interest to clinicians, offering insights into skin irradiation and its associated side effects.

An interesting extension of this work would involve applying it to recent Surface Guided Radiotherapy (SGRT) acquisitions, which generate multiple surface scans per treatment session. Our method could analyze both inter- and intra-fraction variability on this high-quality data, offering valuable insights to improve patient follow-up during radiotherapy.
\section{Acknowledgements}
This work has been supported by the French Ministry of Higher Education and Research through a doctoral scholarship. The data collection protocol during the clinical trial has
been approved by the Ethics committee at the Faculty of
medicine of the University of Strasbourg.

\section{Conflict of interest statement}

The authors declare no conflicts of interest.

\clearpage


\section*{References}
\addcontentsline{toc}{section}{\numberline{}References}
\vspace*{-20mm}





\bibliography{./mybib}      



\bibliographystyle{./medphy.bst}    


\end{document}